\documentclass[hyphens,pdflatex,sn-mathphys-num,iicol]{sn-jnl} 
\usepackage{hyperref}
\hypersetup{
    breaklinks=true,
    colorlinks=true,
    urlcolor=blue,
    linkcolor=blue,
    citecolor=blue
}
\usepackage{breakurl} 

\usepackage{graphicx}%
\usepackage{multirow}%
\usepackage{amsmath,amssymb,amsfonts}%
\usepackage{amsthm}%
\usepackage{mathrsfs}%
\usepackage[title]{appendix}%
\usepackage{xcolor}%
\usepackage{textcomp}%
\usepackage{manyfoot}%
\usepackage{booktabs}%
\usepackage{algorithm}%
\usepackage{algorithmicx}%
\usepackage{algpseudocode}%
\usepackage{listings}%
\usepackage{threeparttable}

\theoremstyle{thmstyleone}%
\theoremstyle{thmstyletwo}%
\theoremstyle{thmstylethree}%

\begin{document}

\title[Article Title]{A Comparative Study of PDF Parsing Tools Across Diverse Document Categories}


\author*[1]{\fnm{Narayan S.} \sur{Adhikari}}\email{n.adhikari2010@gmail.com}

\author[1,2,3]{\fnm{Shradha} \sur{Agarwal}}\email{sabrc@mst.edu}


\affil[1]{\orgdiv{JadooAI}, \orgaddress{\city{Sacramento}, \state{California}, \country{USA}}}

\affil[2]{\orgdiv{Department Nuclear Engineering and Radiation Science},
\orgname{Missouri University of Science and Technology}, \orgaddress{\city{Rolla}, \state{Missouri}, \country{USA}}}

\affil[3]{\orgdiv{Department of Computer Engineering}, \orgname{Missouri University of Science and Technology}, \orgaddress{\city{Rolla}, \state{Missouri}, \country{USA}}}



\abstract{PDF is one of the most prominent data formats, making PDF parsing crucial for diverse NLP tasks, including document classification, information extraction, and retrieval, especially with the growing prevalence of Retrieval Augmented Generation (RAG) framework. 
While various PDF parsing tools exist, their effectiveness across different document types remains understudied, especially beyond academic documents. Our research aims to address this gap by comparing 10 popular PDF parsing tools across 6 document categories using the DocLayNet dataset. These tools include PyPDF, pdfminer.six, PyMuPDF, pdfplumber, pypdfium2, Unstructured, Tabula, Camelot, as well as the deep learning-based tools Nougat and Table Transformer(TATR). We evaluated both text extraction and table detection capabilities. For text extraction, PyMuPDF and pypdfium generally outperformed others, but all parsers struggled with Scientific and Patent documents. For these challenging categories, learning-based tools like Nougat demonstrated superior performance. In table detection, TATR excelled in the Financial, Patent, Law \& Regulations, and Scientific categories. Table detection tool Camelot performed best for Government Tenders, while PyMuPDF performed superiorly in the Manual category. Our findings highlight the importance of selecting appropriate parsing tools based on document type and specific tasks, providing valuable insights for researchers and practitioners working with diverse document sources.}




\maketitle

\section{Introduction}\label{sec1}

PDF(Portable Document Format) was developed in 1992 to enable viewing and exchanging electronic documents independently of device or environment\cite{ISO32000-1:2008}. It uses an imaging model derived from the PostScript language. PDF can incorporate various types of content, including text, images, annotations, videos, and 3D objects. Also, PDF supports encryption, digital signatures, attachments, and metadata. These features have made it one of the most popular document formats. It is estimated that roughly around 2.5 trillion PDF documents are in circulation\cite{Staar2018}.
PDF parsing is crucial for a wide range of NLP tasks, including document summarization, translation, information retrieval, and question answering. With the advent of promising Retrieval-Augmented Generation (RAG) architectures\cite{Lewis2020}, it has become even more important, as PDF is a common source for knowledge base creation, and unlike tagged documents such as HTML, PDFs only store instructions for character and line placement\cite{Lin2024}. Parsing PDFs has several critical challenges that demand careful consideration and handling\cite{Bast2017}, some of these are mentioned below:
\begin{enumerate}
  \item Word identification: Extraction processes may incorrectly break words, mishandle hyphenation, or struggle with special characters like emojis and diacritics (e.g., `a' vs. `à').
  \item Word order preservation: Maintaining correct word sequence can be problematic, especially when dealing with multi-column layouts.
  \item Paragraph integrity: Text flow can be disrupted by embedded formulas or images, potentially fragmenting paragraphs or inadvertently incorporating image captions into the main text.
  \item Table extraction: Inaccurate identification/ complete failure in the identification of tabular data. This can result in misaligned rows and columns, compromising the integrity of the extracted information.
\end{enumerate}

PDF parsing methods can be broadly categorized into rule-based and learning-based approaches. Rule-based methods include fixed rules\cite{Alamoudi2021}, Hidden Markov Models\cite{Hetzner2008}, etc.
Learning-based approaches include a variety of techniques such as using Machine learning\cite{Tkaczyk2015}\cite{Lopez2009}, CRNN\cite{Shi2016}, and transformer architectures\cite{Kim2022}\cite{Li2023}.
While advanced machine learning methods are promising in this area, studying rule-based parsers for PDF analysis remains highly relevant. Rule-based approaches offer distinct advantages in computational efficiency, speed of deployment, and ease of use. They require less processing power and can be quickly implemented without any domain-specific training, making them accessible to a wider range of users and scenarios. One of the other advantages of rule-based parsers is their interpretability, which allows for easier debugging and auditing of the parsed text. 
The primary objective of our study is to evaluate the state-of-the-art rule-based tools for text extraction from PDFs across various domains, including their performance on both general text and tabular content. We aim to identify their shortcomings and propose potential solutions to address these limitations.
\\
We conduct a comprehensive comparison of 10 well-maintained, open-source PDF parsing tools using the DocLayNet dataset\cite{Pfitzmann2022} for general text extraction and table detection tasks. Notably, this is the first comparative study of PDF parsers across six distinct document categories. To the best of our knowledge, the DocLayNet dataset has not been previously utilized for such studies.
Our study primarily utilizes digitally-born PDFs rather than being scanned from paper documents from the DocLayNet dataset. For such documents, rule-based parsers are one of the most efficient methods for text extraction. To contextualize our work, we summarize labeled datasets and evaluation approaches used in previous studies in Section \ref{sec2}. Section \ref{sec3} outlines the DocLayNet dataset and our evaluation criteria, establishing the foundation for our comparative analysis of PDF parsing tools. At the end part of the paper, we provide a comparison of PDF parsers across document categories as well as an overall study, aiming to provide a fair assessment of their capabilities for table and text extraction.

\section{Related Work}\label{sec2}

We first survey the existing labeled datasets for information extraction tasks from PDFs. The earliest datasets for Document Image Analysis and Recognition (DIAR) can be traced back to the 1990s, including NIST\cite{Grother1995} and UW\cite{Phillips1993} datasets. Over the past 30 years, there has been significant progress in this direction. Notably, since 2015, the field has seen a boom in terms of dataset availability. Initially, the datasets were small and comprised scanned PDFs or images, but as time progressed, datasets with digitally born PDFs became more prevalent.
In creating a Document Layout Analysis (DLA) dataset, annotation is the most challenging part. The process of annotation can be broadly classified into three categories\cite{Gemelli2024}:
\begin{enumerate}
  \item Manual: A set of rules is given to human annotators for annotating the documents.
  \item Automatic: A set of algorithms is used to annotate the data. Humans are only needed for quality checking.
  \item Generative: Generative models are used to synthesize the data.

\end{enumerate}
Manual annotation is very laborious and not scalable for large documents. Automatic annotation is a good choice for a large number of documents, but it has certain constraints. It often requires additional structured files such as TeX or XML. 
Most of the DLA datasets consist of scientific or research documents as shown in Table \ref{tab1}. There are two primary reasons for this: (i) Availability: These documents are easily accessible online, e.g., arXiv. (ii) Ease of annotation: Automatic annotation is possible since most of these accompanying TeX files.
\\
We also observe that some of these datasets are partially annotated, focusing only on certain elements (such as metadata or references) of the documents. DocBank and PubLayNet are currently the two largest fully annotated datasets available. PubLayNet\cite{Zhong2019}, containing over 360,000 documents, was constructed using scientific and medical publications. On the other hand, DocBank\cite{li-etal-2020-docbank} was created using approximately 500,000 documents from arXiv. It categorizes the extracted text into 12 element categories. The recent M6Doc\cite{Cheng2023} dataset contains 9,080 manually annotated pages, which include scanned and photographed documents from categories such as scientific articles, textbooks, books, test papers, magazines, newspapers, and notes in Chinese and English. There are many datasets specifically dedicated to table detection and table structure recognition. PubTables-1M\cite{Smock2022} is the largest dataset created using scientific articles for table detection and structure recognition. It has input files in PDF/XML format and output as JSON. However, several popular datasets, such as ICDAR-2019 \cite{Gao2019} and TableBank \cite{li-etal-2020-tablebank}, contain input in Image/LaTeX format. We have not included them in Table \ref{tab1} as we are interested in PDFs only.
\\
DocLayNet\cite{Pfitzmann2022} is the largest dataset containing fully annotated digital-born documents from six different domains (Law and Regulations, Financial documents, Government Tenders, Scientific articles, Manuals, and Patents). It comprises over 80,000 manually annotated documents categorized into 11 different element categories (Caption, Footnote, Formula, List-item, Page footer, Page-header, Picture, Section-header, Table, Text, and Title).

\begin{table*}[ht]
\caption{Overview of commonly cited datasets for information extraction from PDFs, detailing various types of ground truth elements (GTE) including references (R), full text with layout details (FT), and tables (T). The ground truth elements were generated either automatically using XML or LaTeX files, or manually with human intervention.}\label{tab1}%
\begin{tabular}{@{}llllll@{}}
\toprule
\textbf{Dataset} & \textbf{Size} & \textbf{Source} & \textbf{Document Type} & \textbf{GTE} & \textbf{Annotation} \\
\midrule
GIANT \cite{Grennan2019} & 1B & Crossref & Research articles & R & Automatic(XML) \\
S2ORC\cite{Lo2020} & 8.1M & Semantic Scholar & Research articles & R, FT & Automatic(Latex) \\
PubLayNet\cite{Zhong2019} & 360k & PubMed & Biomedical articles & FT & Automatic(XML) \\
SciTSR\cite{Chi2019} & 15k & arXiv & Research articles & T & Automatic(Latex) \\
Bast\cite{Bast2017} & 12k & arXiv & Scientific articles & FT & Automatic(Text) \\
DocBank\cite{li-etal-2020-docbank} & 500k & arXiv & Research articles & FT & Automatic(Text) \\
FinTabNet\cite{GTE} & 89k & Multiple sources & Annual financial reports & T & Automatic(XML) \\
PubTables-1M\cite{Smock2022} & 1M & PubMed & Scientific articles & T & Automatic(XML) \\
DocLayNet\cite{Pfitzmann2022} & 80k & Multiple sources &Multiple & FT & Manual \\
M6Doc\cite{Cheng2023} & 9k & Multiple sources & Multiple & FT & Manual \\
SciBank\cite{Grijalva2022} & 74k & arXiv & Scientific articles & FT & Automatic(Latex) \\
\botrule
\end{tabular}
\end{table*}

Several studies have been conducted to evaluate PDF parsing tools using various metrics (Table \ref{tab2}). Our literature review of these studies found that:
\begin{enumerate}
  \item Some of these studies\cite{Tkaczyk2015}\cite{Lipinski2013} focus solely on selected element extraction capabilities of PDF parsers (e.g. metadata).
  \item Only Bast\cite{Bast2017} and Meuschke\cite{Meuschke2023} have compared PDF parsers for full layout analysis.
  \item Almost all of these studies have tested PDF parsers against academic documents exclusively.
  \item There is no study comparing PDF parsers that focus solely on full-text extraction without specific element extraction (e.g., headers, titles).
  \item Meuschke et al.\cite{Meuschke2023} shows that the Table extraction quality of some of these tools(Camelot, Tabula, etc.) is significantly lower compared to other elements. However, they do not further investigate the underlying reasons for this disparity.
  \item Except for the study by Meuschke\cite{Meuschke2023}, most of the tools from other studies are outdated or not actively maintained.
\end{enumerate}

These observations highlight the need for a comparative study to evaluate the performance of the latest PDF parsers across a wide range of document types, not just limited to scientific publications. The DocLayNet dataset includes various document types with specific element labels such as formulas and tables. The diversity in document categories in the DocLayNet dataset allows for a more accurate representation of the variety of document layouts found in real-world applications. We chose this dataset for our study because it directly aligns with our objectives: (i) comparing PDF parser performance across multiple document types; (ii) assessing parser capabilities for comprehensive full-text extraction; and (iii) evaluating table extraction performance.

\begin{table*}[h]
\caption{Summary of studies comparing PDF parsers. Evaluation metrics used by the studies: Precision(P), Recall(R), and F1 Score.}\label{tab2}%
\begin{tabular}{@{}lccccc@{}}
\toprule
\textbf{Paper} & \textbf{Dataset Size} & \textbf{Document Type} & \textbf{Metrics} & \textbf{Elements} & \textbf{No. of tools} \\
\midrule
Tkaczyk\cite{Tkaczyk2015} & 9,491 & Scientific & P, R, F1 & References & 10 \\
Bast\cite{Bast2017} & 12,000 & Scientific & Custom & Multiple & 14 \\
Lipinski\cite{Lipinski2013} & 1,253 & Scientific & Accuracy & Metadata & 9 \\
Meuschke\cite{Meuschke2023} & 500,000 & Academic & P, R, F1 & Multiple & 10 \\
\botrule
\end{tabular}
\end{table*}

\section{Methodology}\label{sec3}

In this section, we provide an overview of the main features of the dataset we used for our study, namely the DocLayNet dataset\cite{Pfitzmann2022}, and outline the steps we took to generate the ground truth text. Additionally, we discuss the evaluation metrics utilized and the PDF parsers evaluated in our analysis.

\subsection{DocLayNet Dataset}\label{subsec2}
\begin{figure}[h]
\centering
\includegraphics[width=\columnwidth]{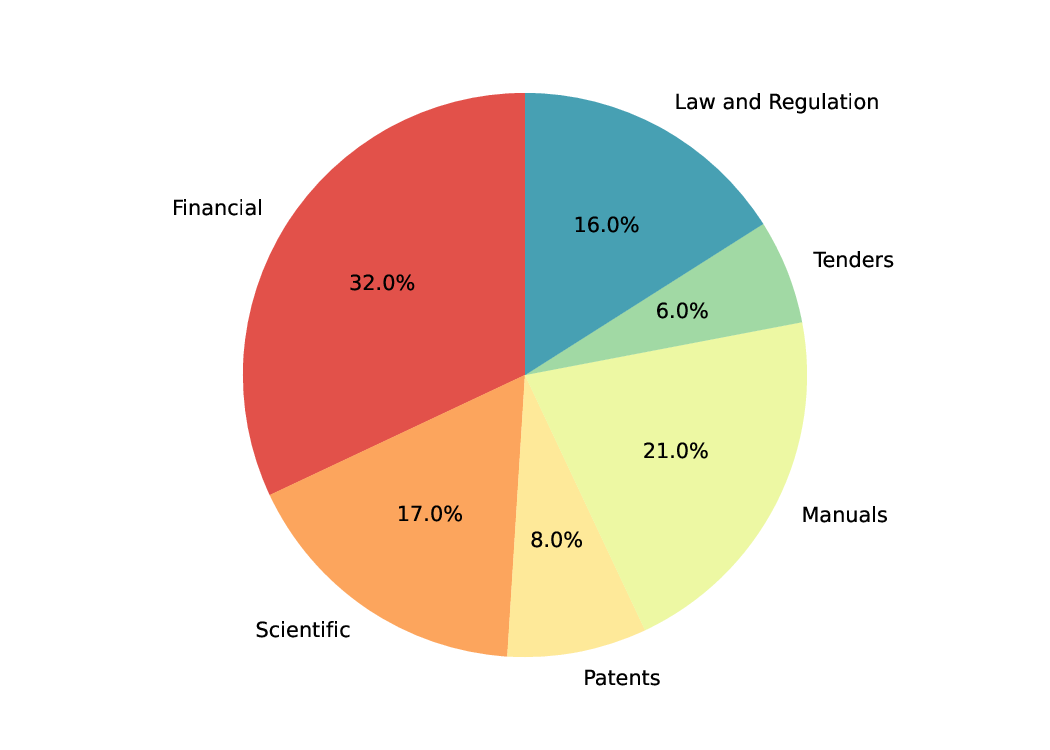}
\caption{Distribution of document categories in DocLaynet Dataset\cite{Pfitzmann2022} }\label{fig1}
\end{figure}

DocLayNet contains approximately 80,000 document pages. Documents are annotated with 11 distinct elements: Footnote, Formula, List-item, Page footer, Page-header, Picture, Section header, Table, Text, and Title. The documents provided in the DocLayNet dataset are classified into 6 distinct categories: Financial Reports, Manuals, Scientific Articles, Laws and Regulations, Patents, and Government Tenders. The distribution of these categories is provided in Figure \ref{fig1}. These documents are mostly in English (95\%), with a few documents in German (2.5\%), French (1\%), and Japanese (1\%).

The other datasets\cite{Zhong2019}\cite{li-etal-2020-docbank}mainly contain scientific documents taken from repositories such as arXiv or PubMed. These datasets have limited variability in layout as they follow more or less uniform templates. However, DocLayNet provides a wide range of document layouts. The `Financial' and `Manual' categories include a large number of freestyle documents. Specifically, Financial Reports consist of both annual reports in freestyle format and formal SEC (Securities and Exchange Commission) filings, while the Manuals category comprises documents such as instructions for computer program manuals and grammar guides. The remaining categories - Scientific Articles, Laws and Regulations, Patents, and Government Tenders - contain documents from various websites and publishers, further increasing the variability in document layouts. To ensure the high quality and reliability of the annotations, around 7,059 documents were doubly annotated, and 1,591 documents were triply annotated. This means these documents were independently annotated by two or three different annotators respectively, allowing for the determination of inter-annotator agreement.
\\
DocLayNet's `core' dataset contains JSON files in standard COCO format\cite{lin2014microsoft} with images (PNG). Each JSON file has information such as document category, document name, precedence (non-zero in case of redundant double- or triple-annotation), bounding box coordinates, and text inside the bounding boxes. DocLayNet's `extra' dataset contains PDF and JSON files which include the text and bounding box coordinates. Both datasets contain files split into test, train, and validation sets. 

\subsubsection{Extraction of Ground Truth}\label{subsubsec1}

For the extraction of ground truth, we used the processed files from Hugging Face \footnote{https://huggingface.co/datasets/pierreguillou/DocLayNet-base}, which contained the DocLayNet core dataset with corresponding PDF files taken from DocLayNet extra files.
For text extraction, we followed a 4-step process to generate ground truth from JSON :

\begin{enumerate}[1.]
\item Load the JSON into a data frame.

\item Sort the text by `id\_box\_line'. The `id\_box\_line' is a unique identifier that ensures the text is processed in the correct order based on its position in the document. 

\item Add the text together with a space if the text `category' is the same. If the text `category' changes, add a new line.

\item Repeated the process for each JSON file.

\end{enumerate}

\begin{figure}[h]
\centering
\includegraphics[width=\columnwidth]{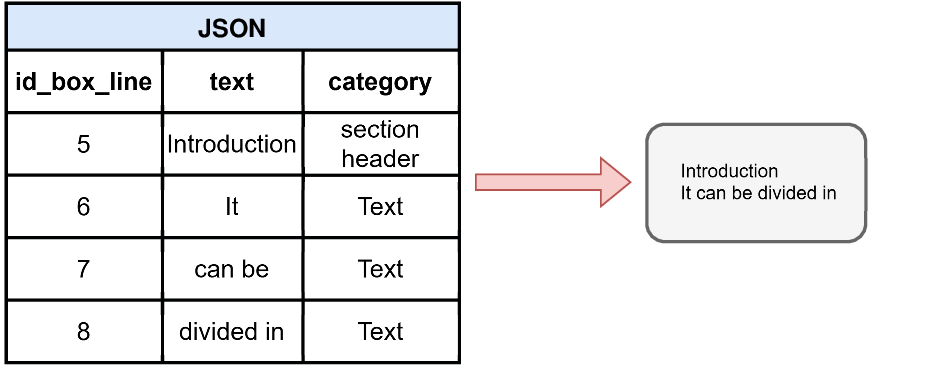}
\caption{Example of ground truth generation from the JSON file loaded into a dataframe. Content from the `text' column was extracted, and new lines and spaces were added according to the `category' column.}\label{fig2}
\end{figure}

With these steps, we were able to extract the full ground truth text as closely as possible to the actual layout. In addition to that we made sure that the header is always on top and the footer on bottom. An example of the output is shown in Figure \ref{fig2}.

\subsection{Evaluation Procedure}\label{subse2}

\begin{figure}[h]
\centering
\includegraphics[width=\columnwidth]{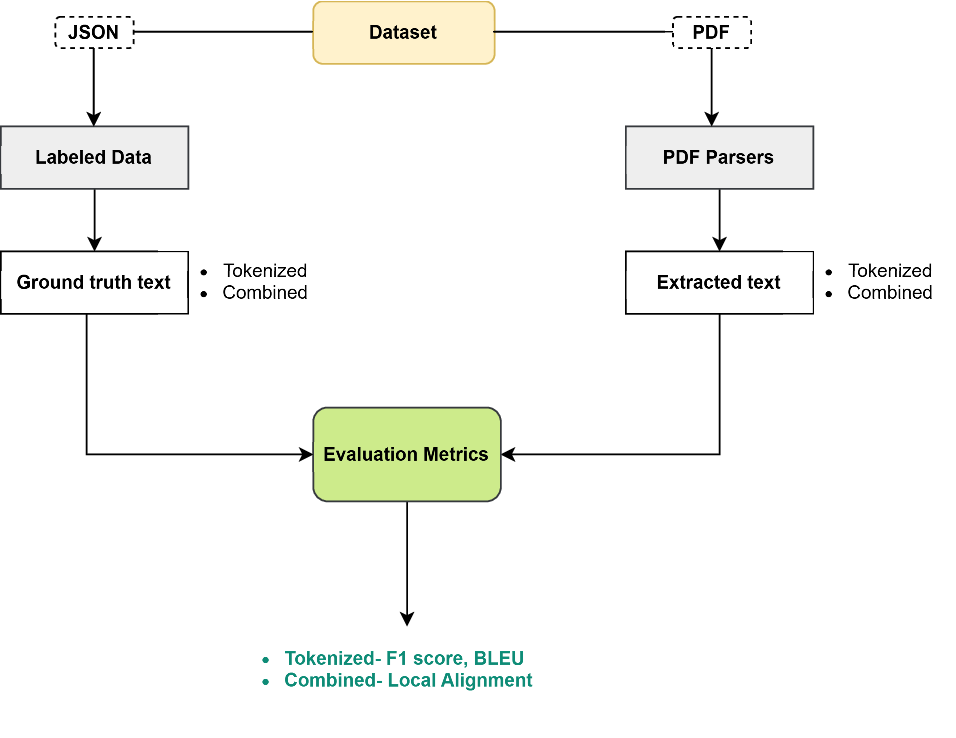}
\caption{Comparison of PDF parser outputs against ground truth data. Both JSON and PDF files are processed to produce ground truth text and extracted text from PDF parsers. Both outputs are saved in the tokenized and combined format before being evaluated using metrics such as F1 score, BLEU, and Local Alignment.}\label{fig3}
\end{figure}

We used two different evaluation procedures: i) for the text and ii) for the tables. The distinction is necessary because, in the former, we are evaluating the text extraction quality of the parser, while in the latter, we are only evaluating the table detection ability of parsers.
\\
\textbf{i) For the text extraction}: Ground truth text was obtained by parsing JSON files using the procedure discussed in the previous section. Correspondingly, PDFs with matching filenames were processed using a PDF parser to obtain the extracted text. We compared the text extracted from PDF parsers with the ground truth from the JSON file. However,  for some metrics(Levenshtein similarity and BLEU) as shown in Figure \ref{fig3}, the text is required in tokenized format, we tokenized the combined text for those metrics.

\textbf{ii)For the table detection}:  We used a similar process as described in the previous section. However, we only extracted `text' from the JSON file if it belonged to the `Table' category. The rule-based PDF parser we used has the capability to extract the tables separately (Table \ref{tab1}). However, these parsers only provide the tables and not the bounding boxes of the identified tables. Therefore, we relied on the extracted text to determine whether the table was correctly detected. We compared the extracted text from the tables recognized by the PDF parsers with the ground truth text and then used a threshold to decide whether the table was correctly identified. When comparing with transformer-based parsers, instead of using the extracted text, we relied on the bounding boxes since they provide the bounding box coordinates in the output. The threshold criterion and the evaluation metrics will be discussed in detail in the subsequent section.

\subsection{Evaluation Metrics}\label{subsec3}
In this section, we establish evaluation criteria suitable for comparing extracted text against the ground truth and for assessing table detection. As discussed in the introduction, several factors can affect extraction quality, including word order, word identification, paragraph alignment, and misidentification of tables. Therefore, it is crucial to utilize evaluation metrics that consider all these factors.
\\

\textbf{For Text Extraction}:
For the evaluation of text extraction, we used a three-fold evaluation strategy:
i) Calculation of F1 score using Levenshtein similarity
ii) BLEU Score
iii) Calculation of local alignment score

The Levenshtein distance($L_d$) is the minimum number of edit operations required to transform one string into another. The edit operations are: (i) Single character insertion (ii) Single character deletion (iii) Single character substitution.
\\
Levenshtein Similarity($L_s$) for two strings $s_1,s_2$ is defined in Equation \ref{eq1}. However, we prefer Normalized Levenshtein similarity defined in Equation \ref{eq2} as it is not sensitive to the length of strings\cite{Tashima}.

\begin{equation}
    L_s(s_1,s_2)=1-L_d(s_1,s_2)
    \label{eq1}
\end{equation}

\begin{equation}
    \|L_s(s1,s2)\| = \frac{L_s(s_1,s_2)}{max(l_1,l_2)}
    \label{eq2}
\end{equation}
where,

$l_1,l_2$ are lengths of strings $s_1,s_2$.

For each document, we generated a similarity matrix by computing the Normalized Levenshtein similarity score between the tokenized extracted text and the ground truth.
Each element of  Similarity Matrix S (Equation \ref{eq3}) represents the similarity between the $i^{th}$ token of extracted text and $j^{th}$ token of ground truth. 
\begin{equation}
    S=[S_{ij}]_{(l_{et} \times l_{gt})}
    \label{eq3}
\end{equation}
We chose a matching threshold of 0.7\cite{Meuschke2023}.
We then computed Precision, Recall, and $F_1$ scores as follows:
\[
  TP_{i,j} =
  \begin{cases}
    \text{1} & \text{if $s_{ij} \geq $ 0.7 }\\
    \text{0} & \text{Otherwise}\\
  \end{cases}
\]
Where,
$s_{ij}$ is an element of the similarity matrix.
\begin{align}
P &= \frac{\mathop{\sum_{i=1}^{l_{et} }\sum_{j=1}^{l_{gt}}} TP_{i,j}}{l_{et}}
\label{eq4} \\
R &= \frac{\mathop{\sum_{i=1}^{l_{et} }\sum_{j=1}^{l_{gt}}} TP_{i,j}}{l_{gt}}
\label{eq5} \\
F_{1}&= \frac{2 \times P \times R}{P+R}
\label{eq6} 
\end{align}
This procedure is discussed in detail by Meuschke et al.\cite{Meuschke2023}. This metric evaluates the PDF parser's word/token-wise extraction quality. By computing the Normalized Levenshtein similarity token-wise, the F1 score provides a reliable estimate of the PDF parser's ability to accurately identify and extract individual words from the PDF.

\begin{figure}[h]
\centering
\includegraphics[width=\columnwidth]{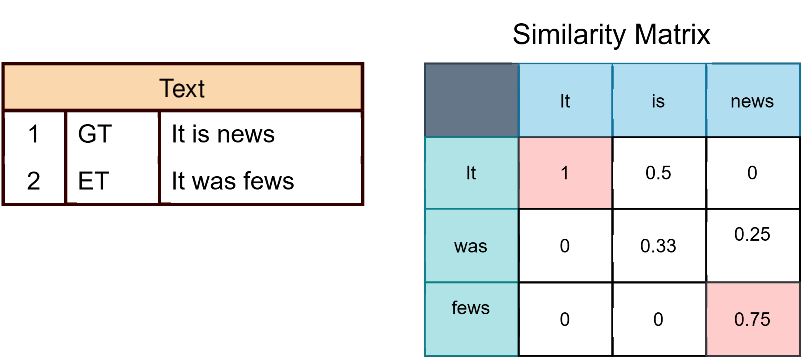}
\caption{Similarity matrix is generated by calculating the normalized Levenshtein similarity between tokenized GT(Ground truth) and ET(Extracted text). if the value is greater than the threshold(colored) it is counted as 1. Here $\sum_{i=1}^{3}\sum_{j=1}^{3} TP_{i,j} =2$ }\label{fig4}
\end{figure}

The BLEU (Bilingual Evaluation Understudy)\cite{Papineni} method was originally developed for evaluating machine translation. It can be used to compare a reference text (the ground truth) with a candidate text (the extracted text). To calculate the BLEU score, we first compute the geometric average of the modified n-gram precision $p_n$ of the tokenized ground truth and extracted text. Then, we multiply it by the brevity penalty (BP),
\begin{equation}
\text{BP} = \begin{cases}
1 & \text{if } c > r \\
\exp(1-r/c) & \text{otherwise.}
\end{cases}
\end{equation}
and,
\begin{equation}
\text{BLEU} = \text{BP} \cdot \exp\left( \sum_{n=1}^N w_n \log p_n\right)
\end{equation}
where c is the length of candidate text(extracted text) and r is the length of reference text(ground truth text). N is the maximum length of n-gram, $w_n$ assigned weights to n-gram precision. In our experiments, we calculated BLEU-4 score(N=4). Since BLEU computes the n-gram overlap between the extracted text and the ground truth, it is an effective metric for evaluating both word order and word identification.

\begin{figure}[h]
\centering
\includegraphics[width=\columnwidth]{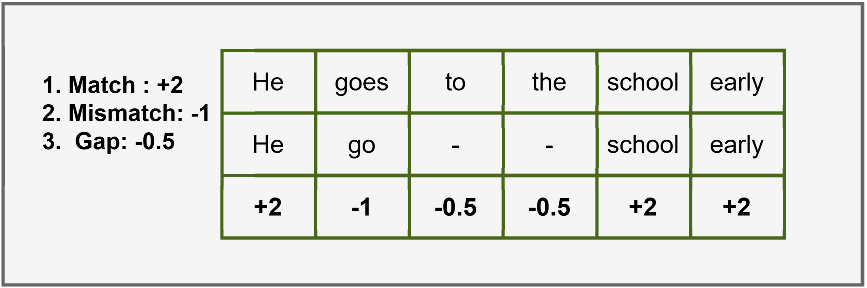}
\caption{An example of Local alignment score calculation for two strings. First, we define the matching score, mismatch, and Gap penalty. For these two strings, the local alignment score is 4 and the normalized local alignment score is 0.67.}\label{fig5}
\end{figure}

Local alignment is a commonly used method in bioinformatics for matching sequences\cite{cock}. We use the local alignment score to assess the overall quality of text extraction. When given two strings $s_1$ and $s_2$, we look for two substrings $s'_{1}$ and $s'_{2}$ (from $s_1$ and $s_2$ respectively) with the highest similarity among all pairs of substrings from $s_1$ and $s_2$. The similarity is calculated using a scoring system (refer to Figure \ref{fig5}), where matches receive positive scores, and mismatches and gaps are penalized. The gap penalty\cite{cock} can be defined in the following way:

\begin{equation}
   \text{ GS = OGS + (n-1)} \times \text{EGS}
\end{equation}
Where,
GS is Gap Score,
OGS is an open gap score for the first gap in a cluster.
EGS is the extended gap score used for each gap following the open gap.
n is the length of the gap.\\
Local alignment score is a quantifiable scoring system, that balances well-matched areas against parsing errors like incorrect word extraction, layout mistakes, and paragraph splitting issues, making it an excellent choice for evaluating PDF parsers. The local alignment score is usually calculated using the Smith-Waterman algorithm\cite{SMITH1981195}.
In our experiments, we calculated the normalized local alignment(normalized by the length of a longer string) score using combined ground truth and extracted text.

\textbf{For Tables}:
In order to evaluate table extraction, we use the Intersection over Union (IoU) to compare the similarity between the table extracted by the parser and the ground truth table. IoU can be defined as follows:

\begin{equation}
    \text{IoU} = \frac{\mid A \cap B \mid}{\mid A \cup B \mid}
    \label{eq10}
\end{equation}
Where A and B can be areas of bounding boxes or sets of strings; the latter is also known as Jaccard similarity.

If the parser extracts text from a table without providing the bounding boxes, we use the Jaccard index to calculate the precision and recall of the detection. First, we flatten the tables extracted by parsers into a list, as the tables from the ground truth JSON can only be extracted as a list. Then, for each document, we compute the normalized Jaccard similarity between all extracted tables and ground truth tables. If the Levenshtein similarity between a pair of ground truth tables and an extracted table exceeds a threshold, we consider it as a correctly identified table by the parser.

In cases where bounding box information is available, we use Intersection over Union (IoU) to calculate the precision and recall of the detection. The DocLayeNet dataset contains bounding boxes in COCO format, but we converted it to Pascal VOC format because the model we used requires this format.\footnote{COCO to PASCAL: $[x_{center}, y_{center}, width, height] \rightarrow [x_{min}, y_{min}, x_{max}, y_{max}]$} Then we computed the IoU between the extracted table and ground truth table according to the Equation \ref{eq10}. If the IoU is greater than the threshold, we consider it a correctly identified table. Finally, we calculate precision, recall, and F1 score for table detection.
\subsection{Tools used}\label{subsec4}
In our study, we conducted a comprehensive comparison of 10 open-source tools for text and table extraction tasks. These tools were selected based on their recent activity, ensuring that each has had active contributions on GitHub within the last six months. This criterion guarantees that the tools are up-to-date and likely to be supported and maintained by their developers.
The tools we evaluated are summarized in Table \ref{tab3}. By including only actively maintained tools, we aim to present the most relevant and effective solutions available for text and table extraction tasks.\\
\textbf{PyPDF}\footnote{https://github.com/py-pdf/pypdf}:  PyPDF is a mature, pure Python library capable of extracting text, images, and metadata from PDF files. It has inspired many forks, including the well-known PyPDF2, PyPDF3, and PyPDF4. Notably, PyPDF2 has been merged back into the main PyPDF library, consolidating its features and improvements. For our experiments, we used the latest version of PyPDF.\\
\textbf{Pdfminer}\footnote{https://github.com/pdfminer/pdfminer.six}: Pdfminer is a versatile tool capable of extracting text, images, table of contents, and font size information from PDF files. It performs automatic layout analysis and supports CJK (Chinese, Japanese, Korean) languages as well as vertical writing. For our experiments, we used its most active fork, pdfminer.six.\\
\textbf{PDFPlumber}\footnote{https://github.com/jsvine/pdfplumber}: Built on top of pdfminer, it can extract text as well as tables.  It also features a visual debugging tool to aid in the extraction process.\\
\textbf{PyMuPDF}\footnote{https://github.com/pymupdf/PyMuPDF}: It provides Python bindings to the MuPDF library written in C. It can extract text, tables, and images, and provides optional OCR support with Tesseract. However, here we use only the rule-based version of PyMuPDF for our analysis.\\
\textbf{Pypdfium2}\footnote{https://github.com/pypdfium2-team/pypdfium2}: Pypdfium2 is a binding to the PDFium library, capable of extracting text and images from PDF files.\\
\textbf{Unstructured}\footnote{https://github.com/Unstructured-IO/unstructured}: Unstructured is a library for preprocessing and ingesting images and text documents. It supports element-wise text extraction and can extract images as well. Unstructured also provides support for the OCR and chipper model, to extract text from scanned documents, and performs layout analysis with the `detectron2' model. It offers table extraction features with OCR. We haven't used the OCR version of this tool in our comparison.\\
\textbf{Camelot}\footnote{https://github.com/camelot-dev/camelot}: It is a Python library that provides table extraction features for PDFs. Tables are extracted as Pandas DataFrames. It provides user flexibility to tweak the configuration parameters for table extraction. It uses two technologies: stream and lattice. Lattice mode identifies the demarcated lines between cells and uses them to parse the tables. On the other hand, stream mode uses whitespace between cells to parse the table. In our experiments, we used the default settings.\\
\textbf{Tabula}\footnote{https://github.com/chezou/tabula-py}: Tabula or tabula-py is a Python wrapper around tabula-java and uses PDFBox in the background. It can extract tables from PDFs and convert them into DataFrames, CSV files, or JSON. It also offers stream and lattice modes. In our experiments, we did not specify any mode ourselves.\\
\textbf{Nougat}: Nougat\cite{blecher2023nougat}(Neural Optical Understanding for Academic documents) is a transformer-based vision and document understanding (VDU) model. It uses an encoder-decoder architecture inspired by the donut model. It is specifically trained for academic documents. Nougat excels at converting Scientific documents to markup text and is particularly adept at parsing Mathematical equations.\\
\textbf{TATR}: The Table Transformer(TATR)\cite{smock2021tabletransformer} is an object detection model trained on the PubTables-1M and FinTabNet datasets for table detection. It is capable of recognizing tables from image inputs. However, a separate OCR model is needed to extract the text. TATR can be trained for other domains using custom datasets.

\begin{table*}[h]
\centering
\caption{Overview of text and table extraction tools used in our study. Key extraction capabilities include extraction of Image (I), Text (T), Metadata (M), Table of Contents (TOC), and Table (TB). Most tools use rule-based (RB) technology, with some offering Optical Character Recognition (OCR) capabilities. However, Nougat and Table Transformers were not the primary focus of this study.}\label{tab3}%
\begin{tabular}{@{}llllr@{}}
\toprule
\textbf{Tool} & \textbf{Version} & \textbf{Extraction} & \textbf{Technology} & \textbf{Output} \\
\hline
PyPDF & 4.3.0 & I, T, M & RB & TXT \\
\hline
pdfminer.six & 20240706 & I, T, TOC & RB & TXT, HTML, hORC, JPG \\
\hline
PyMuPDF & 1.24.7 & I, T, TB & RB(MuPDF), OCR & TXT, HTML, SVG, JSON \\
\hline
pdfplumber & 0.11.2 & I, T, TB & RB(pdfminer) & TXT, HTML, hORC, JPG \\
\hline
pypdfium2 & 4.30.0 & T & RB & TXT \\
\hline
Unstructured & 0.14.10 & T, TB & RB, OCR & TXT \\
\hline
Tabula & 2.9.3 & TB & RB & DataFrame, CSV, JSON \\
\hline
Camelot & 0.11.0 & TB & RB & DataFrame, CSV, JSON, HTML \\
\hline
Nougat & base(350M) & T & Transformer & Markdown \\
\hline
Table Transformer & TATR-v1.1-All & TB & Transformer& Image \\
\hline
\end{tabular}
\end{table*}

\section{Results}
In this section, we present the results of evaluating various PDF parsers for text extraction and table detection tasks using the DocLayNet dataset\cite{Pfitzmann2022}. We compared parsers across 6 categories as summarized in Table \ref{tab4}-\ref{tab5} and Figures \ref{fig6}-\ref{fig11}.
For text extraction(\ref{subsec5}), we used a Levenshtein similarity threshold of 0.7. For table detection(\ref{subsec6}), the Jaccard index threshold was set to 0.75, and we computed IOU with two thresholds: 0.6 and 0.7.
To ensure a fair comparison, we used a balanced dataset with an equal number of documents across all categories in our experiments.

\subsection{For Text Extraction:} \label{subsec5}
\begin{figure}[h]
\centering
\includegraphics[width=\columnwidth]{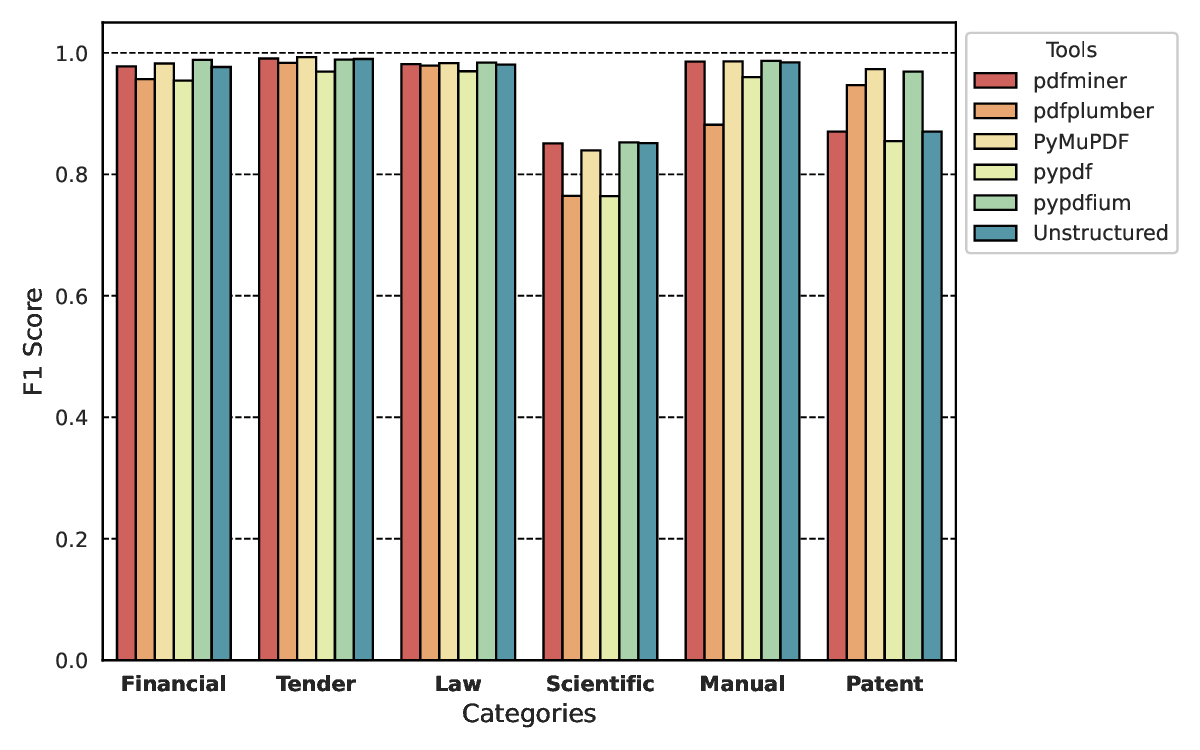}
\caption{F1 score of 6 PDF parsers across all document categories for text extraction. }\label{fig6}
\end{figure}

\begin{figure}[h]
\centering
\includegraphics[width=\columnwidth]{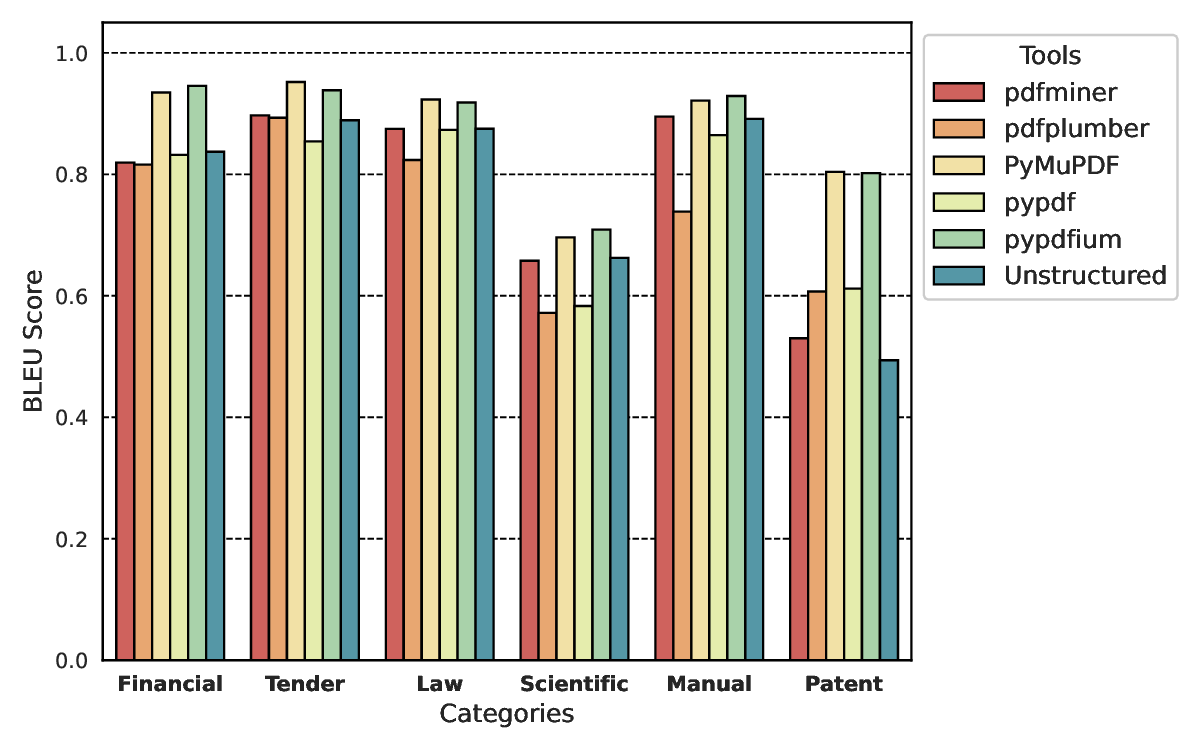}
\caption{BLEU-4 score of 6 PDF parsers across all document categories for text extraction.}\label{fig7}
\end{figure}

We compared the PDF parsers using the metrics:  F1 scores, BLEU-4, and local alignment scores across 6 document categories(refer Table \ref{tab4}).
The results reveal significant performance variations among PDF parsing tools across different document categories. Financial, Tender, Law, and Manual categories saw consistently high F1 scores across most tools, with PyMuPDF and pypdfium consistently performing better in these document types as shown in Figure \ref{fig6} and Figure \ref{fig7}. 
 PyMuPDF and pypdfium demonstrated consistency in word order preservation, achieving the highest BLEU-4 scores in Financial, Manual, Scientific, and Tender, Patent, Law \& Regulations categories, respectively. This suggests that these two tools are particularly adept at maintaining the original word structure in sentences, a crucial factor in many information retrieval applications.
PyMuPDF and pypdfium have also shown good performance in local alignment scores indicating their ability to handle complex layouts and paragraph structures effectively. Additionally, PyPDF has demonstrated high local alignment scores across some categories such as Law (0.9358) and Manual (0.9343).  However, it's important to note that it didn't perform as well in other metrics, emphasizing the significance of considering multiple evaluation criteria when choosing a parser.

\begin{figure}[H]
\centering
\includegraphics[width=\columnwidth]{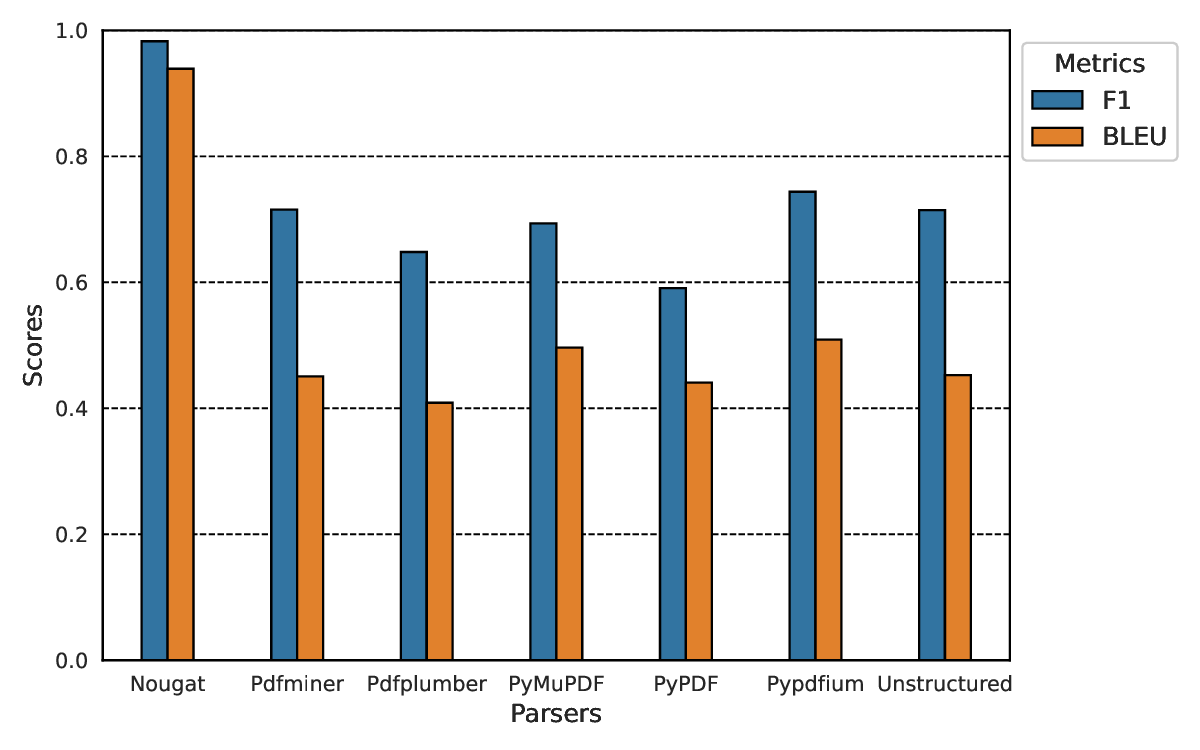}
\caption{Comparison of rule-based parsers and Nougat for text extraction in Scientific documents using F1 and BLEU.}\label{fig8}
\end{figure}

\begin{figure*}[h]
\centering
\includegraphics[width=\textwidth]{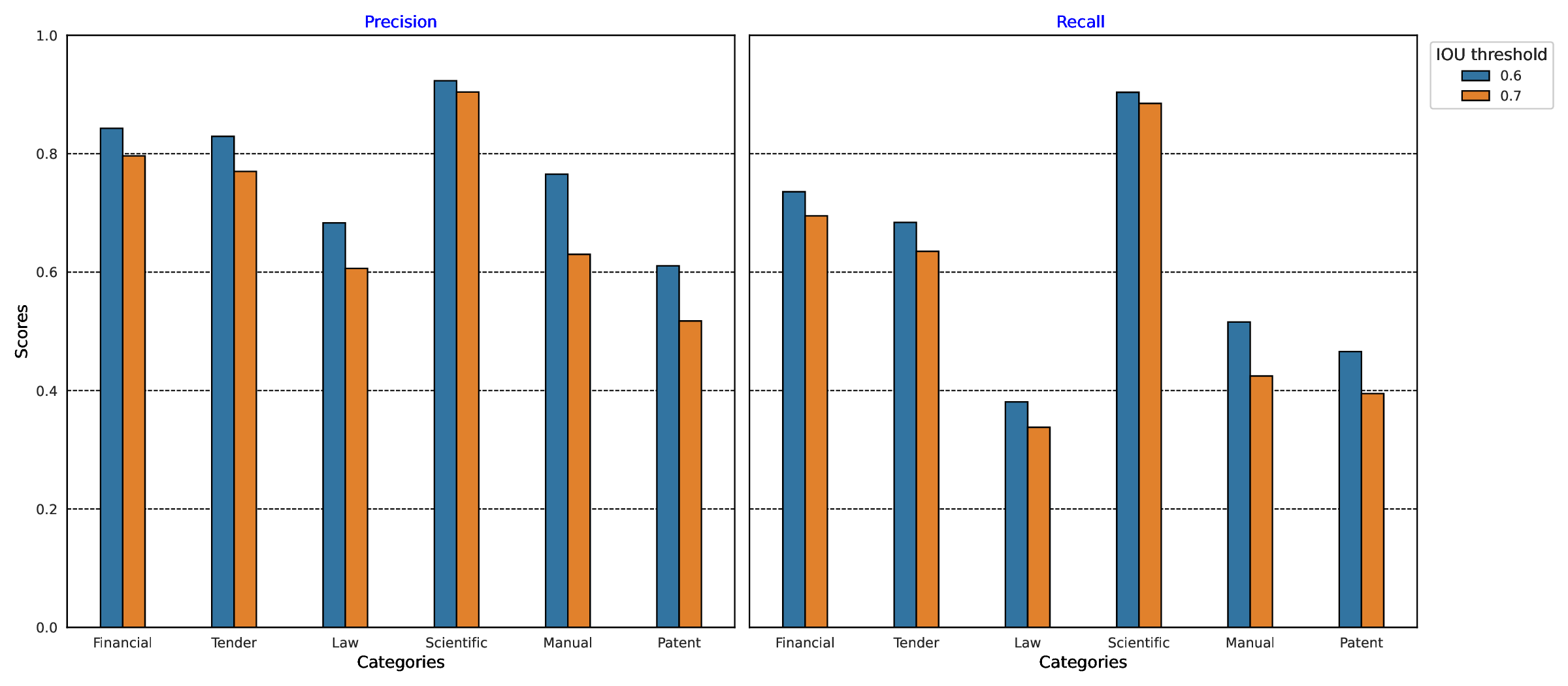}
\caption{Precision and Recall of Table Transformer(TATR) across all document categories for Table detection.}\label{fig11}
\end{figure*}

Scientific and Patent categories presented notable challenges. In the Scientific category, all tools showed a marked decrease in performance, with Pypdfium maintaining a slight edge at an F1 score of 0.8525 and BLEU score of 0.7089, as shown in Table \ref{tab4}.
The Patent category exhibited the widest performance gap among tools. PyMuPDF and pypdfium significantly outperformed others,  scoring F1 scores of 0.973 and 0.969 respectively. 
The significant drop in scores for Scientific and Patent documents highlights a persistent challenge in PDF parsing technology. 
For Scientific documents, we compare the rule-based parser with Nougat a Visual transformer model and it outperforms all rule-based parsers by a huge margin as shown in Figure \ref{fig8}.

The results show that the type of document has a strong influence on the performance of the tools used. The best choice of tools depends on the specific document type and the performance aspects prioritized for a particular information retrieval task. For tasks that prioritize maintaining the structure of the document, like legal document analysis, parsers with high BLEU-4 scores such as PyMuPDF may be preferred. On the other hand, tasks requiring comprehensive information capture may benefit from parsers with high recall. Although there was not a significant difference in F1 scores among the tools (the highest difference being 0.1 for the patent category), the variation in BLEU and local alignment scores was quite apparent as shown in Table \ref{tab4}. This indicates that the differences in performance among parsers primarily lie in their ability to accurately interpret the structure and layout of the documents, rather than in their ability to extract individual words.

\begin{figure}[h]
\centering
\includegraphics[width=\columnwidth]{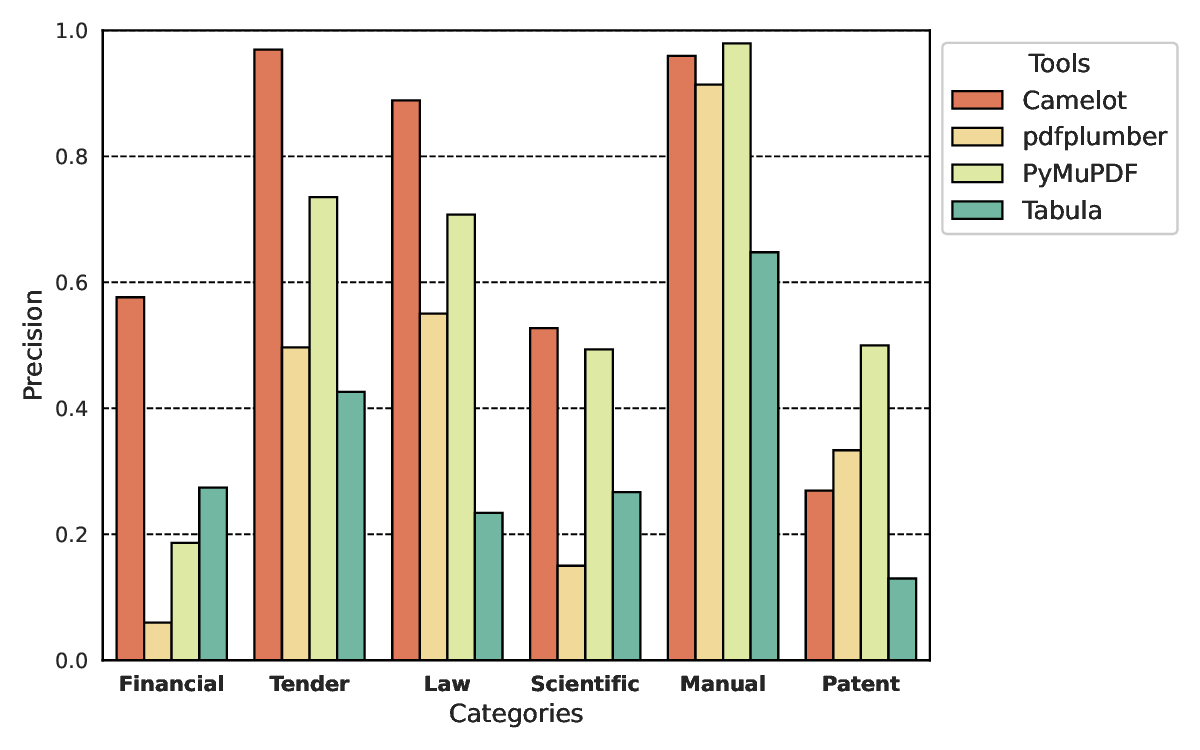}
\caption{Precision of rule-based parsers across all document categories for Table detection.}\label{fig9}
\end{figure}

\begin{figure}[h]
\centering
\includegraphics[width=\columnwidth]{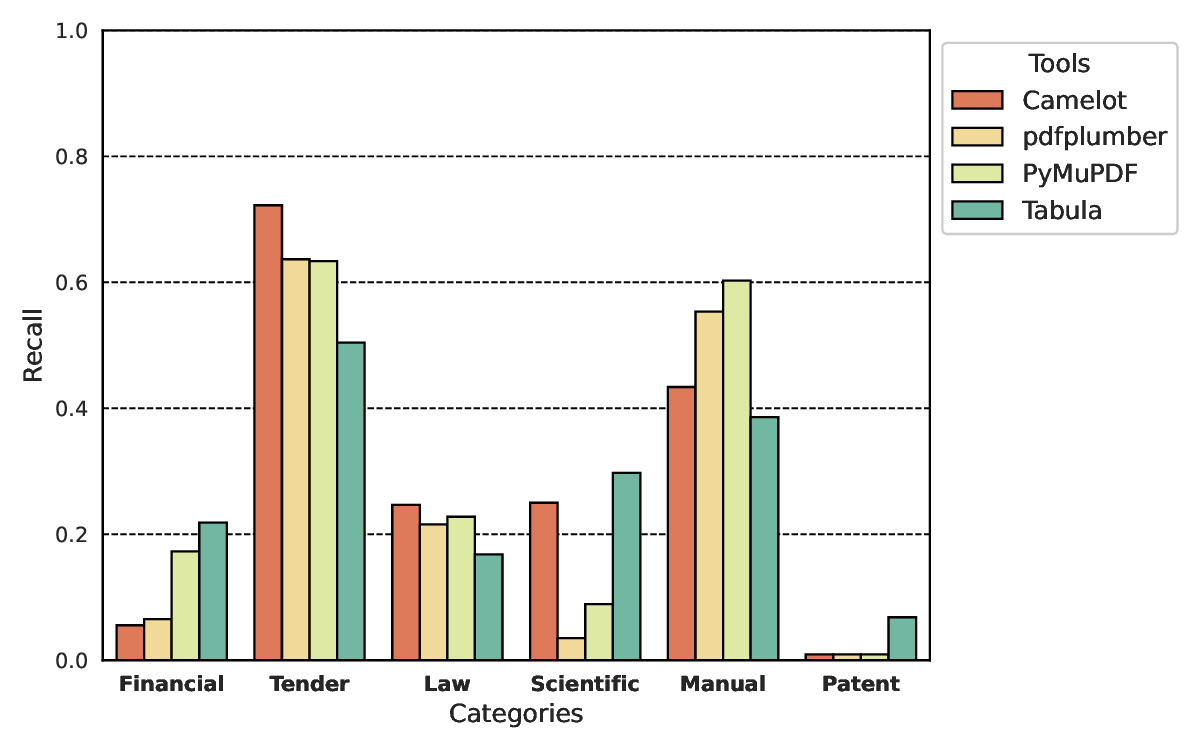}
\caption{Recall of rule-based parsers across all document categories for Table detection.}\label{fig10}
\end{figure}

\begin{table*}[h]
\centering
\caption{A comprehensive comparison of various PDF parsing libraries across different document categories (Financial, Law and Regulations, Manual, Patent, Scientific, and Government tenders). Performance is evaluated using: F1 score, Precision, Recall, BLEU score, and Local Alignment. Higher values (indicated by $\uparrow$) are better for all metrics. Bold values represent the best performance for each metric within each category. The evaluation used 800 balanced documents per category, ensuring a fair comparison across different document types.
}
\small
\begin{tabular}{llccccc}
\hline
\textbf{Category} & \textbf{Parser} & \textbf{F1 ($\uparrow$)} & \textbf{Precision ($\uparrow$)} & \textbf{Recall ($\uparrow$)} & \textbf{BLEU ($\uparrow$)} & \textbf{Local Alignment ($\uparrow$)} \\
\hline
\multirow{6}{*}{Financial} & pdfminer.six & \textbf{0.9979} & 0.9649 & \textbf{0.9912} & 0.8191 & 0.6827 \\
 & pdfplumber & 0.9568 & 0.9785 & 0.9361 & 0.8159 & 0.7029 \\
 & PyMuPDF & 0.9825 & 0.9760 & 0.9892 & 0.9348 & 0.9178 \\
 & pypdf & 0.9542 & 0.9612 & 0.9474 & 0.8321 & 0.8978 \\
 & pypdfium & 0.9885 & \textbf{0.9909} & 0.9860 & \textbf{0.9457} & \textbf{0.9285} \\
 & Unstructured & 0.9767 & 0.9649 & 0.9887 & 0.9371 & 0.8371 \\
\hline
\multirow{6}{*}{Law} & pdfminer.six & 0.9814 & 0.9796 & \textbf{0.9832} & 0.8748 & 0.7996 \\
 & pdfplumber & 0.9791 & 0.9815 & 0.9768 & 0.8236 & 0.6506 \\
 & PyMuPDF & 0.9831 & 0.9857 & 0.9806 & \textbf{0.9232} & 0.9354 \\
 & pypdf & 0.9698 & 0.9746 & 0.9650 & 0.8732 & \textbf{0.9358} \\
 & pypdfium & \textbf{0.9839} & \textbf{0.9912} & 0.9768 & 0.9183 & 0.9228 \\
 & Unstructured & 0.9807 & 0.9798 & 0.9816 & 0.8751 & 0.8359 \\
\hline
\multirow{6}{*}{Manual} & pdfminer.six & 0.9857 & 0.9882 & 0.9832 & 0.8950 & 0.8617 \\
 & pdfplumber & 0.8817 & 0.9672 & 0.8100 & 0.7386 & 0.8432 \\
 & PyMuPDF & 0.9860 & 0.9886 & \textbf{0.9835} & 0.9213 & 0.9317 \\
 & pypdf & 0.9601 & 0.9765 & 0.9442 & 0.8645 & \textbf{0.9343} \\
 & pypdfium & \textbf{0.9868} & \textbf{0.9908} & 0.9829 & \textbf{0.9290} & 0.9311 \\
 & Unstructured & 0.9843 & 0.9893 & 0.9794 & 0.8913 & 0.8835 \\
\hline
\multirow{6}{*}{Patent} & pdfminer.six & 0.8703 & 0.9672 & 0.7910 & 0.5301 & 0.6141 \\
 & pdfplumber & 0.9469 & 0.9538 & 0.9401 & 0.6070 & 0.5459 \\
 & PyMuPDF & \textbf{0.9732} & \textbf{0.9726} & \textbf{0.9737} & \textbf{0.8042} & \textbf{0.8507} \\
 & pypdf & 0.8548 & 0.9291 & 0.7916 & 0.6117 & 0.7842 \\
 & pypdfium & 0.9692 & 0.9709 & 0.9676 & 0.8020 & 0.8108 \\
 & Unstructured & 0.8704 & 0.9672 & 0.7911 & 0.4939 & 0.5873 \\
\hline
\multirow{6}{*}{Scientific} & pdfminer.six & 0.8510 & 0.8918 & \textbf{0.8137} & 0.6577 & 0.7222 \\
 & pdfplumber & 0.7644 & 0.8584 & 0.6890 & 0.5719 & 0.6446 \\
 & PyMuPDF & 0.8395 & 0.8970 & 0.7888 & 0.6962 & \textbf{0.8088} \\
 & pypdf & 0.7641 & 0.8810 & 0.6746 & 0.5832 & 0.7968 \\
 & pypdfium & \textbf{0.8526} & \textbf{0.9046} & 0.8063 & \textbf{0.7089} & 0.8004 \\
 & Unstructured & 0.8514 & 0.8941 & 0.8127 & 0.6625 & 0.7407 \\
\hline
\multirow{6}{*}{Tender} & pdfminer.six & 0.9908 & 0.9915 & 0.9901 & 0.8971 & 0.8333 \\
 & pdfplumber & 0.9834 & 0.9868 & 0.9801 & 0.8932 & 0.8513 \\
 & PyMuPDF & \textbf{0.9929} & \textbf{0.9955} & \textbf{0.9904} & \textbf{0.9521} & \textbf{0.9433} \\
 & pypdf & 0.9691 & 0.9565 & 0.9821 & 0.8544 & 0.9404 \\
 & pypdfium & 0.9888 & 0.9946 & 0.9831 & 0.9385 & 0.9315 \\
 & Unstructured & 0.9899 & 0.9915 & 0.9884 & 0.8890 & 0.8580 \\
\hline
\end{tabular}

\label{tab4}
\end{table*}

\subsection{For Table Extraction:} \label{subsec6}
The evaluation of four rule-based PDF table extraction tools - Camelot, pdfplumber, PyMuPDF, and Tabula - along with a transformer-based model TATR for table detection shows performance patterns across various document categories(Table \ref{tab5}).
While rule-based tools like Camelot excel in specific document types, the transformer-based model demonstrates superior versatility and consistency across all categories.
In terms of recall, the rule-based tools performed poorly in all categories other than Manual and Tender, as shown in Fig. \ref{fig10}. Camelot achieved the highest score in the Tender category (0.72). Tabula outperformed others in the Manual, Scientific, and Patent categories. PyMuPDF showed the most consistent recall across categories among rule-based tools. 
The Table Transformer, however, demonstrated high recall scores across Scientific, Financial, and Tender categories with Scientific documents achieving the highest recall ($>$0.9) (Fig. \ref{fig11}, right panel). However, in the Manual and Tender categories, its performance is not better than PyMuPDF and Camleot respectively(Table \ref{tab5}).

\begin{table*}[h]
\centering
\caption{A comprehensive comparison of various PDF parsers for table detection across different document categories. For Camelot, pdfplumber, PyMuPDF, and Tabula, the Jaccard threshold is 0.75. Intersection over Union (IoU) threshold for TATR @ 60 and TATR @ 70 is 0.60 and 0.70, respectively. The best scores for each category are highlighted in bold. Higher values (indicated by ↑)
are better for all metrics. The evaluation used 400 balanced documents per category, ensuring a fair comparison across different document types.}
\begin{tabular}{llccc}
\hline
\multirow{2}{*}{\textbf{Category}} & \multirow{2}{*}{\textbf{Parser}} & \multicolumn{3}{c}{\textbf{Metrics}} \\

 &  & \textbf{F1 ($\uparrow$)} & \textbf{Precision ($\uparrow$)} & \textbf{Recall ($\uparrow$)} \\
\hline
\multirow{6}{*}{Financial} & Camelot & 0.1012 & 0.5763 & 0.0555 \\
 & pdfplumber & 0.0623 & 0.0596 & 0.6530 \\
 & PyMuPDF & 0.1794 & 0.1863 & 0.1729 \\
 & Tabula & 0.2432 & 0.2740 & 0.2186 \\
 & TATR @ 60 & \textbf{0.7857} & \textbf{0.8430} & \textbf{0.7357} \\
 & TATR @ 70 & 0.7422 & 0.7963 & 0.6949 \\
\hline
\multirow{6}{*}{Law} & Camelot & 0.3861 & 0.8869 & 0.2466 \\
 & pdfplumber & 0.3100 & 0.5502 & 0.2158 \\
 & PyMuPDF & 0.3446 & \textbf{0.7074} & 0.2277 \\
 & Tabula & 0.1956 & 0.2339 & 0.1681 \\
 & TATR @ 60 & \textbf{0.4890} & 0.6831 & \textbf{0.3808} \\
 & TATR @ 70 & 0.4339 & 0.6062 & 0.3379 \\
\hline
\multirow{6}{*}{Manual} & Camelot & 0.5975 & 0.9595 & 0.4338 \\
 & pdfplumber & 0.6895 & 0.9140 & 0.5535 \\
 & PyMuPDF & \textbf{0.7463} & \textbf{0.9794} & \textbf{0.6028} \\
 & Tabula & 0.4837 & 0.6478 & 0.3859 \\
 & TATR @ 60 & 0.6162 & 0.7653 & 0.5157 \\
 & TATR @ 70 & 0.5072 & 0.6300 & 0.4245 \\
\hline
\multirow{6}{*}{Patent} & Camelot & 0.0181 & 0.2692 & 0.0094 \\
 & pdfplumber & 0.0182 & 0.3333 & 0.0094 \\
 & PyMuPDF & 0.0184 & 0.5000 & 0.0094 \\
 & Tabula & 0.0894 & 0.2740 & 0.2186 \\
 & TATR @ 60 & \textbf{0.5285} & \textbf{0.6105} & \textbf{0.4659} \\
 & TATR @ 70 & 0.4480 & 0.5175 & 0.3949 \\
\hline
\multirow{6}{*}{Scientific} & Camelot & 0.3392 & 0.5274 & 0.2500 \\
 & pdfplumber & 0.0623 & 0.0596 & 0.0351 \\
 & PyMuPDF & 0.1794 & 0.1863 & 0.0890 \\
 & Tabula & 0.2432 & 0.2740 & 0.2974 \\
 & TATR @ 60 & \textbf{0.9134} & \textbf{0.9233} & \textbf{0.9038} \\
 & TATR @ 70 & 0.8944 & 0.9041 & 0.8850 \\
\hline
\multirow{6}{*}{Tender} & Camelot & \textbf{0.8279} & \textbf{0.9696} & \textbf{0.7224} \\
 & pdfplumber & 0.5580 & 0.4967 & 0.6366 \\
 & PyMuPDF & 0.6808 & 0.7353 & 0.6388 \\
 & Tabula & 0.4619 & 0.4262 & 0.5042 \\
 & TATR @ 60 & 0.7496 & 0.8293 & 0.6939 \\
 & TATR @ 70 & 0.6961 & 0.7700 & 0.6351 \\
\hline
\end{tabular}
\label{tab5}
\end{table*}

\section{Discussion}
For text extraction, all rule-based parsers underperformed in the Scientific and Patent categories. Scientific documents are challenging to parse due to the Mathematical expressions in them. All of the rule-based parsers analyzed here extract the mathematical equation in symbolic form. We found that it is insufficient to express the complex formulas involving vectors, matrices, etc. Also, the parsers sometimes extract the content of graphs and it gets mixed in paragraphs which further worsens the quality of extraction. 
In our opinion, the better way would be to use parsers that can extract the texts from scientific documents as Latex, Markdown, or MathML formats. As we show with an example of Nougat, these approaches are much better for scientific documents than the conventional rule-based approaches.
In the DocLayNet dataset, most patents are documents filled with images of designs or diagrams of chemical compounds, etc. Parsing such documents is beyond the scope of rule-based parsers. Hence, an OCR-based approach would be more suitable for such documents.

Table detection is a challenging part for all of the parsers. Rule-based parsers excel in detecting the table if i) there are clear boundaries in tables and ii) Spaces between columns' text are fixed. However, most of the parsers in this category were not able to parse the PDF at all. It can be seen in low recall/high false positive scores. However, the diversity in table structures across document categories poses a considerable obstacle for rule-based parsers. We found that all of these parsers mainly struggle to detect the tables if i) Tables are nested ii) There is no table boundary iii) the Table is in the form of the table of contents iv) Multiple tables on a single page v) Tables using color differentiation instead of lines vi) Table columns are separated by ``..” or `` - -”.
We found that the learning-based approach would be more suitable for handling complex tables. As can be seen in Table \ref{tab5} - TATR although trained for financial and scientific documents, also excels in other categories.

\section{Conclusion and Future Work}
While our study provides valuable insights into the performance of various PDF parsing tools across different document categories, it has a few limitations - 
The sample size for comparison of Nougat is relatively small due to the nature of the ground truth of the DocLayNet dataset(Ground truth is not in LaTex format).
One major challenge was dealing with scientific documents, which often contain inline equations, which are Mathematical expressions embedded within the text and crucial for the document's integrity. However, the dataset we used didn't support extracting inline equations, making accurate text comparison difficult. Also, comparison to learning-based methods was limited to a single tool each for scientific text extraction (Nougat) and table detection (TATR).

In future studies, it would be valuable to investigate how different types of learning methods perform when trained on samples from these 6 categories. In particular, We plan to test TATR and other such models on a wider range of document types, including Law and Regulation, Manual, Government Tenders, and Patents categories. This will help us understand how well it handles different kinds of tables, like nested or multimodal ones, tables of contents, etc.
During our experiments, we observed that Nougat sometimes hallucinates when parsing Scientific documents. To address this, we would like to explore combining rule-based parsers with learning-based models like Nougat. This hybrid approach might help reduce errors and lead to more accurate parsing. 

In our study, we analyzed and compared the performance of rule-based parsers in extracting text and detecting tables across various document types. We used the DocLayNet dataset and compared these tools using multiple metrics. Our findings show that the performance of these tools is strongly related to the document type or the document structure, with scientific documents presenting the most significant challenges. Based on our findings, we suggest that learning-based approaches could be more suitable for handling scientific documents. 
We also investigated the table detection capabilities of these parsers and found that their poor text extraction performance was largely attributable to inadequate table detection. Based on our findings, we suggest that techniques like TATR could prove more effective for table detection tasks.\\
Additionally, we examined and discussed the underlying factors contributing to the parsers' subpar performance in Scientific categories and table detection. This comprehensive analysis provides valuable insights into the strengths and limitations of current rule-based parsing tools, paving the way for future improvements in document analysis and information extraction technologies.

\bibliography{sn-bibliography}

\end{document}